\documentclass[aps,prb,amsmath,amssymb,showpacs,letterpaper,twocolumn]{revtex4}%
\usepackage{hyperref}
\usepackage{graphicx}
\usepackage{graphicx}%
\usepackage{amsfonts}%
\usepackage{color}
\usepackage{textcomp}

\begin{document}

\title{Simulation of a model nanopore sensor: ion competition underlies device behavior} 
\author{Eszter M\'adai$^{1}$, M\'{o}nika Valisk\'{o}$^{1}$, Andr\'as Dallos$^{1}$, and Dezs\H{o} Boda$^{1,2}$\footnote{Author for correspondence: boda@almos.vein.hu}}
\affiliation{$^{1}$Department of Physical Chemistry, University of Pannonia, P.O. Box 158, H-8201 Veszpr\'em, Hungary}
\affiliation{$^{2}$Institute of Advanced Studies K\H{o}szeg (iASK), Chernel st. 14. H-9730 K\H{o}szeg, Hungary}
\date{\today}


\begin{abstract}
We study a model nanopore sensor with which a very low concentration of analyte molecules can be detected on the basis of the selective binding of the analyte molecules to the binding sites on the pore wall.
The bound analyte ions partially replace the current-carrier cations in a thermodynamic competition.
This competition depends both on the properties of the nanopore and the concentrations of the competing ions (through their chemical potentials).
The output signal given by the device is the current reduction caused by the presence of the analyte ions.
The concentration of the analyte ions can be determined through calibration curves.
We model the binding site with the square-well potential and the electrolyte as charged hard spheres in an implicit background solvent.
We study the system with a hybrid method in which we compute the ion flux with the Nernst-Planck (NP) equation coupled with the Local Equilibrium Monte Carlo (LEMC) simulation technique.
The resulting NP+LEMC method is able to handle both strong ionic correlations inside the pore (including finite size of ions) and bulk concentrations as low as micromolar.
We analyze the effect of bulk ion concentrations, pore parameters, binding site parameters, electrolyte properties, and voltage on the behavior of the device. 
\end{abstract}


\maketitle


\section{Introduction}
\label{sec:intro}

Nanopores can be used as molecular sensors due to their small dimensions comparable to molecular scales.
Biological pores called ion channels have well-defined atomic structures with high specificity.
Synthetic nanopores, on the other hand, have enhanced mechanical and thermal stability.
Also, their geometry (diameter, shape, length) and surface chemistry can be manipulated with continuously improving fabrication techniques in order to suit the desired sensing application.  

The rapidly advancing experimental techniques open the routes to a variety of practical sensing applications including ions, peptides, proteins, organic polymers, triglycerides, and small molecules
\cite{patolsky_mattoday_2005,dekker_nnt_2007,sexton_mbs_2007,grieshaber_sensors_2008,kasianowicz_arac_2008,gyurcsanyi_trac_2008,howorka_siwy_chapt11_2009,howorka_csr_2009,piruska_csr_2010,venkatesan_nnt_2011,liu_csr_2012,miles_csr_2013,turner_csr_2013,ying_analyst_2014,makra_ecc_2014,haywood_ac_2015,liu_mca_2015,arjmanditash_csr_2016,zhang_nanotoday_2016,shi_ac_2017}.
Xu et al.\ \cite{xu_ac_2016}, for example, presented a dual-signal-output (fluorescence and ionic current) nanopore system that could detect Hg$^{2+}$ by plugging the nanopores. 
Sun et al.\ \cite{sun_cc_2015} designed a label-free nanochannel for enantioselective recognition of arginine by addition of chiral selectors.
Ali et al.\ \cite{ali_jacs_2011} constructed a nanobiosensor based on the immobilization of metal-ligand complexes inside polyethylene terephthalate (PET) nanopores for the selective recognition of lactoferrin.
The reliable detection of glycerolipids, fatty acids, and ions in low concentrations is of great importance during the production of bio-origin engine fuels with high hydrogen content in their molecular structures \cite{hancsok_tc_2011}. 

The small dimensions of nanopores (the Debye length being smaller than the pore radius) and the importance of local interactions make it necessary to use molecular-level modeling and accompanying computational techniques to gain additional insight into the molecular mechanisms, predict device functions, and design new devices.
This paper provides a modeling study using reduced representations of electrolyte, nanopore, and membrane. 
In our previous paper \cite{hato-pccp-19-17816-2017} we studied rectifying bipolar nanopores and showed that reduced models work if they capture the overall device physics corretly, namely, if those degrees of freedom that are necessary to reproduce device function are built into the model.

In this work, device function is a measurable and reproducible change in the current of background ions through the nanopore modulated by the analyte molecules that are present in the bath in a given concentration.
The analyte molecule (usually a charged biomolecule) is detected if it interacts with the nanopore selectively.
This interaction can be physical adsorption or chemical binding or something in between in strength.
Often, the active site binding the analyte molecule is carried by another molecule that, in turn, is attached to the nanopore wall using chemical methods.  

If the binding is a reversible process involving physical forces that are weaker than covalent bonds, the probability that the analyte ion binds to the nanopore is proportional to its concentration (via its chemical potential). 
The current reduction, therefore, is also related to the analyte concentration, which, in turn, can be determined from calibration curves (ion current reduction vs.\ analyte concentration).

Weak physical binding forces also make it possible to have a large number of events where analyte ions bind to the pore, but unbind easily due to thermal motion. 
The large number of these events, on the other hand, results in a statistical ensemble of properly sampled states.
These kinds of experimental approaches are named `ensemble methods' in the literature to distinguish them from `single-molecule methods' where advanced spatial and temporal resolution of the instrumentation makes it possible to detect single-molecule events \cite{bayley_cr_2000,gyurcsanyi_trac_2008,makra_ecc_2014}.  
In the resistive pulse sensing technique \cite{kasianowicz_arac_2008}, for example, the frequency of the pulses is related to the bulk concentration of the analyte, while the amplitude and shape of the pulses are rather related to the size and geometry of the particles.

The thermodynamic route of our study is aimed to model the `ensemble methods' by assuming the formation of a statistical ensemble composed of a large amount of binding/unbinding events sampled by a Monte Carlo (MC) method.
The binding interactions are usually very specific that are different in different cases (steric interactions facilitating van de Waals forces and hydrogen bonds, for example).
In this paper, however, our purpose is to examine a general model with just a handful of well-defined tunable parameters and to study the effect of changing these parameters with the goal of understanding the molecular mechanisms that underlie device function.
Therefore, we use the square-well (SW) potential that has two parameters with which the range and strength of the potential can be tuned. 

The magnitude of current reduction is determined by the competition between the analyte ions and the ions carrying the current (potassium, for example) for space around the binding sites inside the pore. 
In this competition, all the parameters used in the reduced model of this study play an important role.
Changing any of these parameters shifts the outcome of the competition and modifies the device's behavior as a sensor.
We provide a systematic study about how the response given by the device depends on the model parameters.

This ion competition makes our study closely related to our earlier works for ionic channels \cite{2000_nonner_bj_1976,gillespie-jpcb-109-15598-2005,boda-jcp-125-034901-2006,boda-bj-93-1960-2007,gillespie-bj-95-2658-2008,gillespie_bj_2008,boda-jgp-133-497-2009,dirk-janhavi-mike,malasics-bba-1798-2013-2010,boda-jcp-134-055102-2011,boda-jcp-139-055103-2013,boda-arcc-2014,gillespie_bj_2014}, where the selective binding of the competing ions (Na$^{+}$ vs.\ Ca$^{2+}$ in calcium channels, for example) underlies the selectivity properties of various ion channels.
Properties of ions (charge and size) and their interactions with the channels are basic determinants of the competition.
This study adds the short range potential selectively binding the analyte ion to the picture, thus making the pore able to function as a sensor.
The binding interaction provides an energetic advantage \cite{gillespie_bj_2008,boda-jcp-134-055102-2011} for the analyte ions in the competition with the background ions, so they can partially replace the background ions in the pore even if they are present in the bath in much smaller concentrations.
Therefore, they can inhibit the current of background ions thus producing a detectable signal.

We show that the model nanopore sensor based on this competition principle can detect an analyte ion present at very low concentrations ($10^{-6}-10^{-3}$ M).
Our analysis shows that the most important drivers of signal strength and sensitivity are the relative concentrations of the analyte and background ions, strength and number of the binding sites, and pore radius.

\section{Model}
\label{sec:model}

The electrolyte model is the ``primitive'' model (a historical name, as this model, though simple, is anything but primitive\cite{boda-arcc-2014,hato-pccp-19-17816-2017}) that represents ions as charged hard spheres of valences, $z_{i}$, and radii, $R_{i}$ ($i$ refers to an ionic species):
\begin{equation}
u_{ij}(r) =
\left\lbrace 
\begin{array}{ll}
\infty & \quad \mathrm{for} \quad r<R_{i}+R_{j} \\
 \dfrac{e^{2}}{4\pi\epsilon_{0}\epsilon} \dfrac{z_{i}z_{j}}{r} & \quad \mathrm{for} \quad r \geq R_{i}+R_{j}\\
\end{array}
\right. 
\label{eq:uij}
\end{equation} 
where $e$ is the elementary charge, $\epsilon_{0}$ is the permittivity of vacuum, $\epsilon$ is the dielectric constant of the solvent, and $r$ is the distance between two ions.
The analyte molecule is also an ion in our study (with $z_{\mathrm{X}}$ valence) so the potential in Eq.\ \ref{eq:uij} acts between the ions of the electrolyte and the analyte ion.  
Using ions of finite size is of crucial importance because the competition of the analyte ions with ions carrying the current is a central phenomenon behind the suppression of the current by the presence of the analyte.

The solvent is represented as a continuum background characterized by two response functions. 
One is the dielectric constant, $\epsilon=78.5$, that describes the screening effect of the water molecules (this is assumed to be the same throughout the system).
The other is a diffusion coefficient function, $D_{i}(z)$, that describes the ability of water molecules to affect the diffusion of ions ($z$ is the coordinate along the main axis of the pore, perpendicular to the membrane).
This function is space dependent in our case; it is a piecewise constant function that is different inside the pore ($R_{i}^{\mathrm{pore}}$)
and in the bulk ($R_{i}^{\mathrm{bulk}}$).
The bulk value is experimental (see Table \ref{tab:pauling}), while $R_{i}^{\mathrm{pore}}$ just scales the current without influencing the $I/I_{0}$ ratio.
Our earlier study \cite{hato-pccp-19-17816-2017} where we compared to explicit-water molecular dynamics simulations for a bipolar nanopore indicated that diffusion coefficients are smaller inside the pore than in the bulk.
Following our other study \cite{matejczyk-jcp-146-124125-2017} here we set the relation $R_{i}^{\mathrm{pore}}=0.1R_{i}^{\mathrm{bulk}}$.

\begin{figure}[t]
\includegraphics*[width=0.45\textwidth]{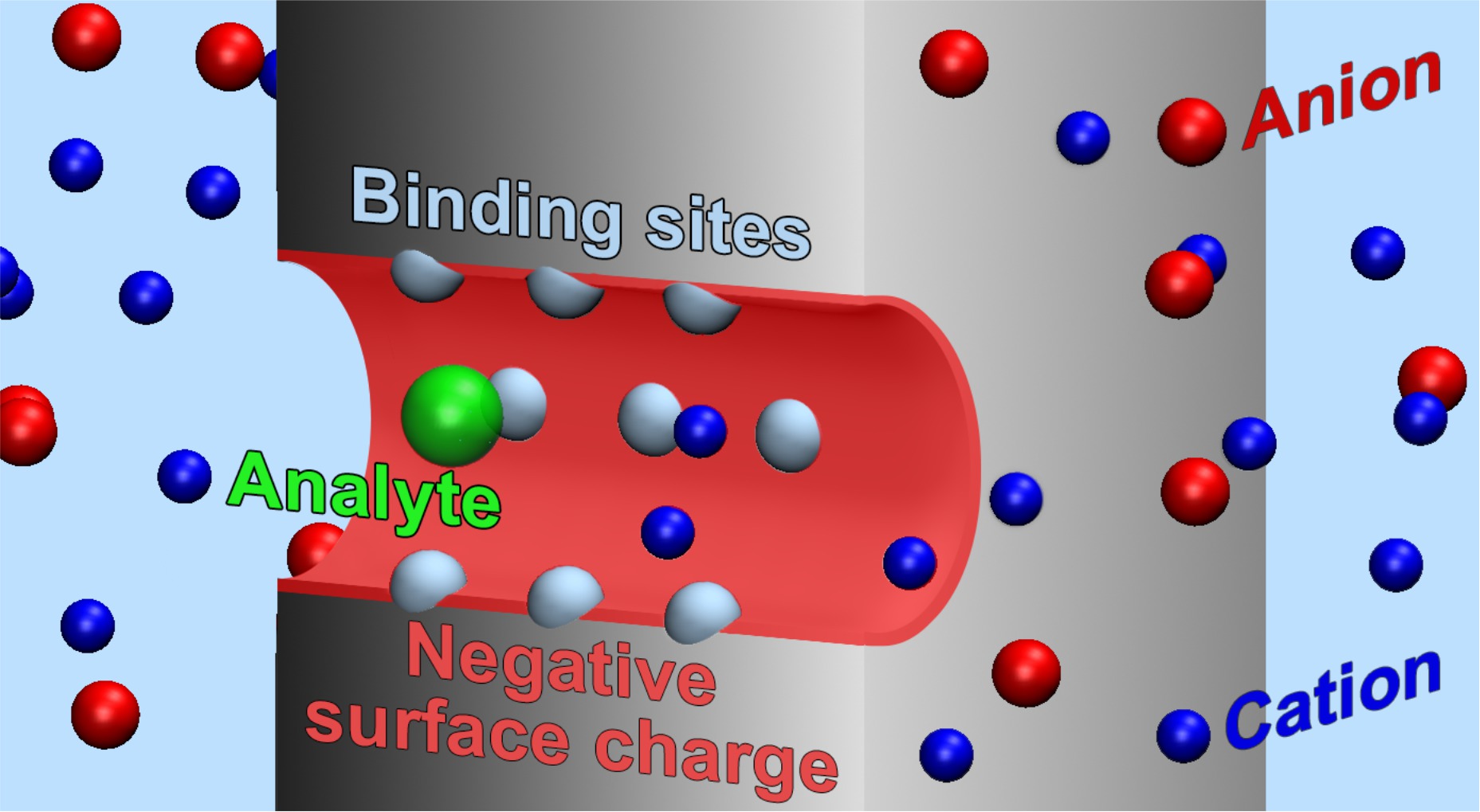}
\caption{
\small (color online) 
Schematics of the pore with grey spheres representing the binding sites (the radius of the sphere equals the $d_{\mathrm{SW}}$ distance parameter of the SW potential), the green sphere is the analyte molecule, X$^{+}$, that is bound to the binding site if it overlaps with the grey sphere. The blue and red spheres are the cations and anions of the electrolyte, respectively. The light red wall of the pore indicates that it carries a negative surface charge.
}
\label{Fig1}
\end{figure} 

The nanopore is a cylindrical pore of radius $R_{\mathrm{pore}}$ penetrating a membrane of width 6 nm.
Note that real nanopores are longer.
This short portion can be viewed as the active segment of the pore carrying binding sites (the tip of a conical nanopore, for example).
The walls of the pore and the membrane are hard, namely, overlap of ions with these walls is forbidden.
We placed the binding sites on the pore wall in $N_{\mathrm{ring}}$ rings; each ring containing 4 binding sites as shown in Fig.\ \ref{Fig1}.
The distance between two rings is 1 nm.
The discrete binding sites break the rotational symmetry of our model, but using 4 of them reduced this effect. 
The binding sites are far enough from each other so that they can be considered independent.
Their exact location, therefore, is not really important compared to their number (for a pore of fixed geometry).
 
The binding potential between a site and an analyte ion is the square-well (SW) potential:
\begin{equation}
u_{\mathrm{SW}}(r) =
\left\lbrace 
\begin{array}{ll}
0 & \quad \mathrm{for} \quad r-R_{\mathrm{X}}>d_{\mathrm{SW}}\\
 -\epsilon_{\mathrm{SW}} & \quad \mathrm{for} \quad r-R_{\mathrm{X}}<d_{\mathrm{SW}} ,
\end{array}
\right. 
\label{eq:squarewell}
\end{equation}
where $r$ is the distance of the site and the ion center and $R_{\mathrm{X}}$ is the radius of X.
This short-range potential attracts X with $-\epsilon_{\mathrm{SW}}$ energy once the closest point of the X ion's surface is closer to the site than the distance parameter $d_{\mathrm{SW}}$.
This model takes into account that the active site of the X ion is usually on its surface while keeping the spherical symmetry of the ion (it neglects the possible orientation dependence of binding).
The SW potential only acts on the X ions in the simulations. 
This is equivalent with having $\epsilon_{\mathrm{SW}}=0$ for all other ion species.
  
Finally, we place negative point charges on the pore wall with which the ions can interact through the Coulomb potential (see Eq.\ \ref{eq:uij}).
They are situated on a rectangular grid, where a surface element is approximately a square of size $0.2 \times 0.2$ nm$^{2}$.
The magnitude of the point charges is established so that the surface charge density corresponds to a prescribed value, $\sigma$.
The purpose of the pore surface charge is to make the pore cation selective. 
In this case, the sensor works more efficiently as shown by the results provided that the analyte ion is also a cation.

\section{Method}
\label{sec:method}

Here, we use a recently developed technique \cite{boda-jctc-8-824-2012}, where the Nernst-Planck (NP) equation is used to compute the ionic flux for ion species $i$
\begin{equation}
 \mathbf{j}_{i}(\mathbf{r})=-\dfrac{1}{kT}D_{i}(\mathbf{r})c_{i}(\mathbf{r}) \nabla \mu_{i}(\mathbf{r}) ,
\label{eq:np}
\end{equation} 
where $T=298.15$ K is temperature, $k$ is Boltzmann's constant, $D_{i}(\mathbf{r})$ is the diffusion coefficient profile of species $i$, $c_{i}(\mathbf{r})$ is the concentration profile, $\mu_{i}(\mathbf{r})$ is the electrochemical potential profile, and $\mathbf{j}_{i}(\mathbf{r})$ is the particle flux density.

We assume that exchange of momentum between ions in negligible so coupling of fluxes can occur only indirectly through molecular interactions between ions.
To handle those interactions and to establish the relation between $c_{i}(\mathbf{r})$ and $\mu_{i}(\mathbf{r})$
we use the Local Equilibrium Monte Carlo (LEMC) simulation method that is an adaptation of the grand canonical Monte Carlo technique to a non-equilibrium situation.
Due to its grand canonical nature, LEMC can handle small concentrations easily.

We divide the computation domain of the NP system into volume elements and use different $\mu_{i}^{\alpha}$ values in each volume element $\mathcal{D}^{\alpha}$.
Insertions/deletions of ions are attempted into/from these volume elements with equal probability.
These trials are accepted or refused on the basis of the Metropolis algorithm \cite{metropolis}.
The acceptance probability contains the local electrochemical potential, $\mu_{i}^{\alpha}$, and the energy change.
The energy includes every interaction from the whole simulation cell, not only from subvolume $\mathcal{D}^{\alpha}$.
The result of the LEMC simulation in the $[n]$th iteration is the concentration in every volume element, $c_{i}^{\alpha}[n]$.

The whole system is solved in an iterative way until conservation of mass ($\nabla \cdot \mathbf{j}_{i}(\mathbf{r})=0$) is satisfied.
The procedure can be summarized as
\begin{equation}
\mu_{i}[n] \,\,\, \xrightarrow{\mathrm{LEMC}}  \,\,\,  c_{i}[n] \,\,\,  \xrightarrow{\mathrm{NP}} \,\,\,  \mathbf{j}_{i}[n] \,\,\, 
\xrightarrow{\nabla \cdot \mathbf{j}=0} 
\,\,\, \mu_{i}[n+1] .
\label{eq:circle}
\end{equation} 
The electrochemical potentials for the next iteration, $\mu_{i}^{\alpha}[n+1]$, are computed from the results of the previous iteration, $c_{i}^{\alpha}[n]$, on the basis of the integral form of the continuity equation:
\begin{equation}
 0 = \int_{\mathcal{D}^{\alpha}}\nabla\cdot\mathbf{j}_{i}(\mathbf{r})\, dV = \oint_{\mathcal{S}^{\alpha}} \mathbf{j}_{i}(\mathbf{r}) \cdot \mathbf{n}(\mathbf{r})\,da,
 \label{eq:cont2}
\end{equation}
where volume $\mathcal{D}^{\alpha}$ is bounded by surface $\mathcal{S}^{\alpha}$ and $\mathbf{n}(\mathbf{r})$ denotes the normal vector pointing outward at position $\mathbf{r}$ of the surface. 
Substituting the NP equation for $\mathbf{j}_{i}(\mathbf{r})$, we obtain
\begin{equation}
0 = \oint_{\mathcal{S}^{\alpha}} D_{i} c_{i}[n] \nabla \mu_{i}[n+1]  \cdot \mathbf{n}\,da 
\end{equation}
for every $\alpha$.
This forms a system of linear equations whose solution provides $\mu_{i}^{\alpha}[n+1]$ for the next iteration.

Because the LEMC always computes $c_{i}[n]$ with a statistical uncertainty, the procedure does not converges to a single value, but the resulting $\mu_{i}[n+1]$ values fluctuate around the solution that is obtained as a running average over iterations.
A mixing procedure is used to improve convergence \cite{boda-jml-189-100-2014}.
The current carried by species $i$ is obtained by integrating over the cross section of the pore 
\begin{equation}
 I_{i}=z_{i}e\,2\pi \int_0^{R_{\mathrm{pore}}} r\, \mathbf{j}_{i}(z,r) \cdot \mathbf{n}_{z}\,dr .
\end{equation}

The efficiency and usefulness of the method (called NP+LEMC) was proved in several publications for ion channels \cite{boda-jml-189-100-2014,boda-arcc-2014,hato-cmp-19-13802-2016} and nanopores \cite{matejczyk-jcp-146-124125-2017,hato-pccp-19-17816-2017}.
The method is designed to compute steady-state ion transport.
The boundary conditions (concentrations and electrostatic potentials) on the two sides of the membrane, therefore, are fixed, so the driving force of the transport is maintained. 

Both components of the methodology are important to study the sensor model efficiently.
Although an approximation, using the NP equation allows one to describe the dynamics of ions with a limited computational effort.
In dynamical simulations (such as Brownian Dynamics that is the obvious choice to simulate the drift-diffusion of ions in an implicit solvent), sampling of ions crossing a narrow pore arriving from a low concentration bath is problematic.

The LEMC component of our simulations, on the other hand, is very efficient in handling ionic correlations (including steric effects) and complex models for the nanopore, the membrane, and the binding site.
The widely used Poisson-Nernst-Planck (PNP) theory cannot handle these correlations, because it uses the Poisson-Boltzmann theory (instead of LEMC) to describe the $c_{i}(\mathbf{r})$ vs.\ $\mu_{i}(\mathbf{r})$ relationship, so it considers ions as point charges interacting with the mean electric field.

It is also possible to couple the NP equation with density functional theory \cite{gillespie_jpcm_2002,gillespie_pre_2003,gillespie-jpm-17-6609-2005,valisko-jpcc-111-15575-2007} to include ionic correlations.
Gillespie successfully applied this method for a one-dimensional model of the RyR Ca$^{2+}$-release ion channel \cite{gillespie-jpcb-109-15598-2005,gillespie_bj_2008,dirk-janhavi-mike,gillespie_bj_2014}. 
A three-dimensional version of Gillespie's model was also introduced and simulated by the NP+LEMC method \cite{boda-arcc-2014}.

\begin{table}[t]
\caption{Pauling radii and bulk phase diffusion coefficients for ions used in this work. 
}
\label{tab:pauling}
\renewcommand{\arraystretch}{0.75}
 \begin{center}
  \begin{tabular}{l| c c}
  Ion & $R_{i}$/nm & $D_{i}^{\mathrm{bulk}}$/$10^{-9}$m$^{2}$s$^{-1}$\\\hline
  Cl$^{-}$ & 0.1805 & 2.032\\
  Li$^{+}$ & 0.06 & 1.029\\
  Na$^{+}$ & 0.095 & 1.334\\
  K$^{+}$ & 0.133 & 1.849\\
  Cs$^{+}$ & 0.169 & 2.056\\ 
  X & 0.1-0.5 & 2.032
  \end{tabular}
  \end{center}
\end{table}

\section{Results}
\label{sec:results}

Our purpose is to study the behavior of our model sensor as a function of various model parameters.
Sensor signal is defined by the degree to which the analyte, at concentration $c$, inhibits the current through the pore ($I$) compared to the current in the absence of the analyte ($I_{0}$).
We will plot the relative current reduction, $1-I/I_{0}$, to characterize sensor signal. 
The larger this signal is, the more measurable the effect of the analyte ion is.
A more refined descriptor is the sensitivity, defined as the slope of the relative current, $I/I_{0}$, as a function of concentration.
The faster $I/I_{0}$ decreases (or $1-I/I_{0}$ increases), the more sensitive is the device. 

We want to cover a wide parameter range but in an efficient way.
Therefore, we define two base points with respect to which we vary the parameters.
The parameters of the base points are $\epsilon_{\mathrm{SW}}=10kT$, $d_{\mathrm{SW}}=0.2$ nm, $N_{\mathrm{ring}}=3$, $R_{\mathrm{pore}}=1$ nm, $R_{\mathrm{X}}=0.3$ nm, $z_{\mathrm{X}}=1$, and $\sigma=-1$ $e$/nm$^{2}$.
The $\epsilon_{\mathrm{SW}}=10kT$ energy of the binding potential (corresponding to 24.77 kJ/mol) is in the ballpark of hydrogen bonds \cite{scheiner_jbc_2001}.
The voltage (the difference of the electrostatic potentials between the right and left hand sides; fixed at 200 mV) is the only driving force of current; ionic concentrations are the same on the two sides of the membrane in the bulks.
The background electrolyte is a 0.01 M KCl ($z_{+}=1$, $z_{-}=-1$).
The Pauling radii are used for the ions.
They are listed in Table \ref{tab:pauling}, together with the diffusion coefficients. 
The two base point are defined with two different analyte concentrations $c_{\mathrm{X}}=10^{-5}$ and $10^{-4}$ M.

\subsection{Importance of cation selectivity}
\label{sec:sel}

First, we analyze the effect of pore surface charge.
A positively charged pore is not feasible because it would repel the positive analyte ions, X$^{+}$.
When the surface charge is negative, however, the pore is more efficient in detecting X$^{+}$. 
Furthermore, an optimal behavior in term of the relative current reduction ($1-I/I_{0}$) is observed. 
In the competition between ions for space in the pore, all the different kinds of effects present in the system influence the outcome: the bulk concentration, the electrostatic interaction of ions with the pore charge, the electrostatic interaction of ions with each other, hard sphere exclusion, and the binding potential of X$^{+}$.   

\begin{figure}[t]
\includegraphics*[width=0.45\textwidth]{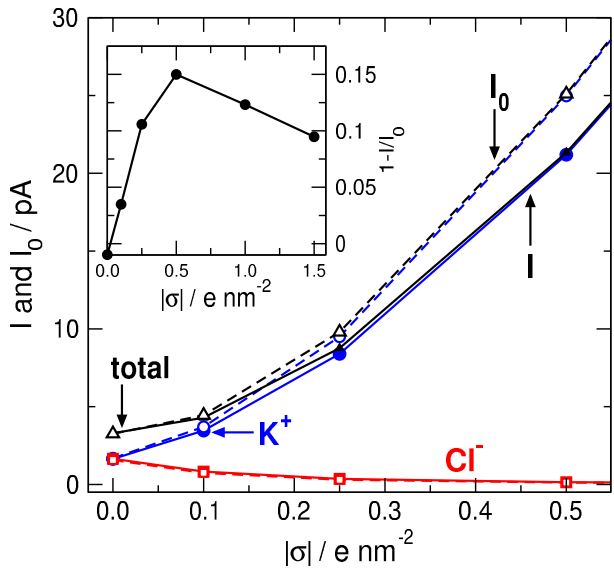}
\caption{
\small (color online)  
Ionic currents (K$^{+}$, Cl$^{-}$ and total) for $c_{\mathrm{X}}=0$ M ($I_{0}$) and $c_{\mathrm{X}}=10^{-5}$ M ($I$) as functions of the magnitude of the pore surface charge, $|\sigma|$ ($\sigma\le0$ in this figure).
Other model parameters are those of the base points as described in the main text.
The statistical errors in currents are usually below 1 \% here and in the figures that follow (error bars are smaller than symbols).
The inset shows the relative current reduction, $1-I/I_{0}$, as a function of $|\sigma|$.
}
\label{Fig2}
\end{figure}

For negative surface charges, Figure \ref{Fig2} shows the currents of K$^{+}$ and Cl$^{-}$ and the total current as functions of the magnitude of the pore surface charge, $|\sigma|$, both in the absence and in the presence of X$^{+}$ ($I_{0}$ and $I$, respectively).
Dashed lines with open symbols and solid lines with filled symbols show the $I_{0}$ and $I$ currents, respectively.  

For zero and very small negative surface charges both K$^{+}$ and Cl$^{-}$ ions enter the pore.
While the X$^{+}$ ions decrease the current of K$^{+}$ ions (due to repulsion), they increase the current of Cl$^{-}$ ions (due to attraction) and there is little change in the total current due to this cancellation (solid and dashed lines go together and $1-I/I_{0}\approx 0$).
An additional disadvantage of this case is that the signal (total current) is small. 

Increasing $|\sigma|$ (namely, making the pore more negative) makes the pore cation selective so the X$^{+}$ ions now compete only with K$^{+}$ ions for space in the pore.
As $|\sigma|$ increases, however, a stronger competition is forced between these two ionic species.
The negative surface charge attracts both K$^{+}$ and X$^{+}$ ions, but the energetic advantage given by the SW potential enables the X$^{+}$ ions to compete efficiently with the K$^{+}$ ions.
The $I$ and $I_{0}$ (solid vs.\ dashed) curves, therefore, increasingly deviate with increasing $|\sigma|$.
Accordingly, the $1-I/I_{0}$ curve increases at moderate $|\sigma|$ (see inset of Fig.\ \ref{Fig2}.)

As $|\sigma|$ is increased further, however, more K$^{+}$ ions will be present in the pore.
The large density of K$^{+}$ ions makes it difficult for X$^{+}$ ions to find space in the pore despite of the SW energetic advantage of the X$^{+}$ ions. 
The $1-I/I_{0}$ curve, therefore, decreases  with increasing $|\sigma|$.

The maximum is at about $\sigma=-0.5$ $e$/nm$^{2}$.
Understanding this behavior more deeply would require an energetic analysis similar to those performed for calcium channels \cite{gillespie_bj_2008,boda-jcp-134-055102-2011,boda-jcp-139-055103-2013,gillespie_bj_2014}, sodium channels \cite{boda-cmp-18-13601-2015}, and double layers \cite{valisko-jpcc-111-15575-2007}.

From now, we use the $\sigma=-1$ $e$/nm$^{2}$ value, because it is generally used in PET nanopores for usual pH values.

\subsection{Dependence on concentrations}
\label{sec:res-conc}

A key result of this paper is the analysis with respect to the concentrations of both the analyte, $c_{\mathrm{X}}$, and the background electrolyte, $c_{\mathrm{K}^{+}}$. 
The probability of an ion entering the pore is proportional to $\ln c_{i}$ that is the ideal part of the chemical potential dominating in the bulk.
The probability of an ion staying inside the pore for a longer period of time, on the other hand, is also affected by the excess part of the chemical potential that, in turn, includes all the interactions of the ion (with other ionic charges, with the binding sites, with the pore charges, exclusion by hard pore wall and by other hard sphere ionic cores).
The product of the two probabilities determines local concentration inside the pore.
Both bulk concentrations and local forces in the pore are important.
The competition of all these effects determines which ion (X$^{+}$ or K$^{+}$) will be present in the pore in larger concentrations; that is, they determine the behavior of the nanopore as a sensor.

\begin{figure}[b]
\includegraphics*[width=0.45\textwidth]{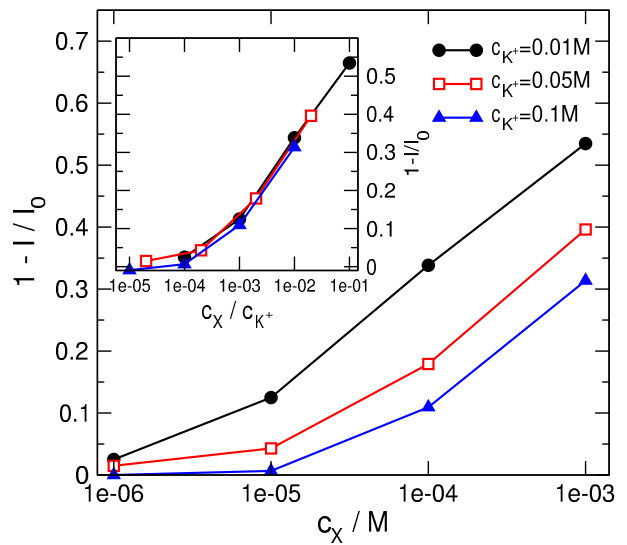}
\caption{
\small (color online)  
Calibration curves: relative current reduction, $1-I/I_{0}$, is plotted, where $I/I_{0}$ is the current relative to the current in the absence of X$^{+}$ ($I_{0}=61.62$, 90.68, and 105.29 pA for $c_{\mathrm{K}^{+}}=0.01$, 0.05, and 0.1 M, respectively) as functions of $c_{\mathrm{X}}$ for different values of $c_{\mathrm{K}^{+}}$. 
Other model parameters are those of the base points as described in the main text.
The inset shows the same data, but as a function of the X$^{+}$ concentration relative to the K$^{+}$ concentration ($c_{\mathrm{X}}/c_{\mathrm{K}^{+}}$).
}
\label{Fig3}
\end{figure} 

Figure \ref{Fig3} shows the current reduction, $1-I/I_{0}$, as a function of $c_{\mathrm{X}}$ for different values of $c_{\mathrm{K}^{+}}$.
The results show that better efficiency can be achieved using lower K$^{+}$ concentration: the black curve increases more steeply than the blue curve, for example.
Because these calibration curves are monotonic, this also means that the black curves are above the blue curves.
Larger $1-I/I_{0}$ values, therefore, tend to correspond to larger slopes, so sensor signal (relative current reduction) and sensitivity (the slope) can be used interchangeably. 
The underlying mechanism is that the X$^{+}$ ion in a given bulk concentration competes more efficiently with K$^{+}$ if K$^{+}$ arrives from a bulk with lower concentration simply because the probability of K$^{+}$ ions trying to enter the pore is lower.

The inset plots $1-I/I_{0}$ as a function of the normalized X$^{+}$ concentration, $c_{\mathrm{X}}/c_{\mathrm{K}^{+}}$.
It shows that the curves coincide, indicating a scalability of the pore's behavior with $c_{\mathrm{X}}/c_{\mathrm{K}^{+}}$.
In practical terms, if we want to detect an X$^{+}$ analyte of a given concentration, we  can find the proper K$^{+}$ concentration that provides the desired signal (the $1-I/I_{0}$ value) on the basis of only one calibration curve ($1-I/I_{0}$ vs.\ $c_{\mathrm{X}}/c_{\mathrm{K}^{+}}$).

We must keep in mind, however, that using very low concentrations can result in smaller currents (the values of $I_{0}$ are shown in the caption of Fig.\ \ref{Fig3}) and larger noise. 
While our methodology cannot account for the noise, the role of signal-noise relation cannot be left unconsidered.
However, as a rule of thumb, one should use a background electrolyte concentration as close to the anticipated analyte concentration as possible.

\begin{figure}[t]
\includegraphics*[width=0.45\textwidth]{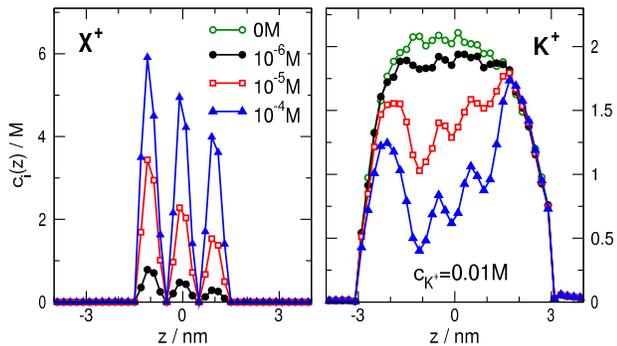}
\caption{
\small (color online) Concentration profiles of X$^{+}$ (left panel) and K$^{+}$ (right panel) ions for various  $c_{\mathrm{X}}$ (see the legends in the left panel) with $c_{\mathrm{K}^{+}}=0.01$ M. 
The model parameters are the base parameters as described in the caption of Fig.\ \ref{Fig3}.
The profiles are obtained by averaging over the cross section of the pore accessible to ions.
}
\label{Fig4}
\end{figure}

The concentration profiles behind the behavior depicted in Fig.\ \ref{Fig3} are shown in Fig.\ \ref{Fig4}.
As $c_{\mathrm{X}}$ is increased in the bulk, the peaks of X$^{+}$ get higher at the $z$ coordinates where the rings of binding sites are located.
Because the X$^{+}$ ions are depleted outside these binding regions due to their very small bulk concentrations, they do not contribute considerably to the total current; the cation of the background electrolyte remains the main charge carrier.
At the same time, the concentration of K$^{+}$ in the pore decreases, causing the drop in the current.

\subsection{Dependence on pore properties}
\label{sec:res-pore}

\begin{figure}[t]
\includegraphics*[width=0.45\textwidth]{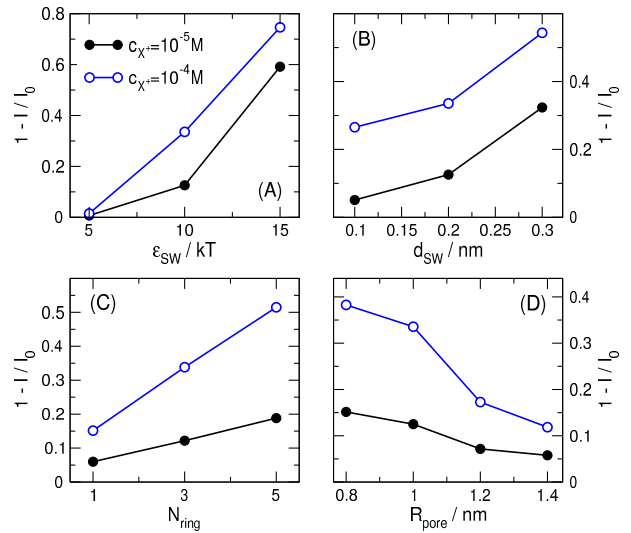}
\caption{
\small (color online) Dependence of the relative current reduction on model parameters (A) $\epsilon_{\mathrm{SW}}$, (B) $d_{\mathrm{SW}}$, (C) $N_{\mathrm{ring}}$, and (D) $R_{\mathrm{pore}}$ for two concentrations of X$^{+}$ ($c_{\mathrm{X}}=10^{-5}$ and $10^{-4}$ M) and $c_{\mathrm{K}^{+}}=0.01$ M. 
Other model parameters are those of the base points as described in the main text.
}
\label{Fig5}
\end{figure}

Next, we analyze the effect of pore and binding-site parameters.
Although $d_{\mathrm{SW}}$ and $\epsilon_{\mathrm{SW}}$ can be viewed as parameters of a pair potential (depending both on the bound and the binding molecule), we choose to consider them as pore properties generated during the fabrication process and discuss their effects here in Fig.\ \ref{Fig5}. 

The distance parameter of the SW potential, $d_{\mathrm{SW}}$, determines the probability that we try to insert X$^{+}$ ions into the regions where the SW attraction acts.
Larger $d_{\mathrm{SW}}$ values make the binding of more X$^{+}$ ions possible, therefore, they are more efficient in decreasing the K$^{+}$ current (Fig.\ \ref{Fig5}B).

The energy parameter, $\epsilon_{\mathrm{SW}}$, on the other hand, influences the probability of accepting the insertion of the X$^{+}$ ion. 
A larger $\epsilon_{\mathrm{SW}}$ value produces a deeper potential well, therefore, the probability of acceptance increases, and so does the concentration $c_{\mathrm{X}}(\mathbf{r})$.
Larger $\epsilon_{\mathrm{SW}}$ values, therefore, promote the efficiency of the sensor (the $1-I/I_{0}$ value increases with increasing $\epsilon_{\mathrm{SW}}$, Fig.\ \ref{Fig5}A).

The same is true for the number of binding sites that we measure in the number of rings, $N_{\mathrm{ring}}$, in Fig.\ \ref{Fig5}C. 
More binding sites can bind more X$^{+}$ that can reduce the K$^{+}$ current more efficiently.

Decreasing the pore radius, $R_{\mathrm{pore}}$, the $1-I/I_{0}$ current reduction increases, namely, the pore becomes more sensitive to the presence of X$^{+}$ ions.
The X$^{+}$ ions occupy a short-range neighborhood of the SW center.
The size of this region depends on the $d_{\mathrm{SW}}$ parameter and independent of the pore radius.
As pore radius is decreased, therefore, the X$^{+}$ ions occupy more space relative to the cross section so they are more effective in the competition with the K$^{+}$ ions (Fig.\ \ref{Fig5}D).

\subsection{Dependence on experimentally controllable parameters}
\label{sec:res-exp}

\begin{figure}[t]
\includegraphics*[width=0.45\textwidth]{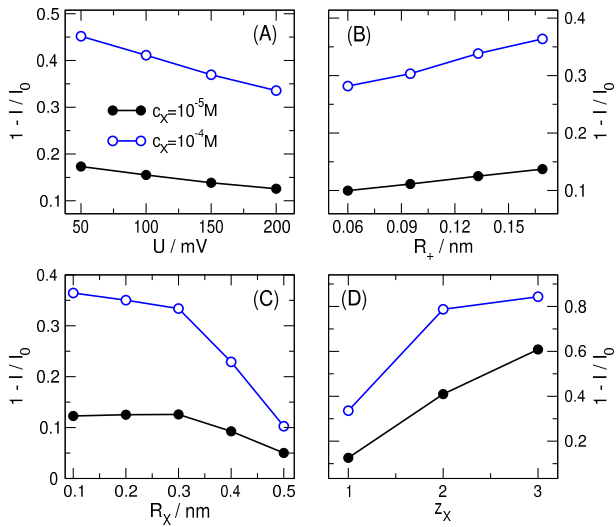}
\caption{
\small (color online) Dependence of the relative current reduction on external parameters (A) $U$, (B) $z_{\mathrm{X}}$, (C) $R_{\mathrm{X}}$, and (D) $R_{+}$ for two concentrations of X$^{+}$ ($c_{\mathrm{X}}=10^{-5}$ and $10^{-4}$ M) and $c_{\mathrm{K}^{+}}=0.01$ M. 
Other model parameters are those of the base points as described in the main text.
}
\label{Fig6}
\end{figure}

While these parameters are characteristic of the nanopore and hard to change after fabrication, there are also parameters that can be controlled easily in an experiment.
The effects of these are analyzed in Fig.\ \ref{Fig6}.
Voltage, for example, is a basic parameter that  can be varied easily.
Large voltages drive K$^{+}$ through the pore more forcibly, so the ability of X$^{+}$ ions to prevent their flow is more limited.
Small voltages, therefore, are advantageous from the point of view of current reduction (Fig.\ \ref{Fig6}A).
They can, however, be disadvantageous from the point of view of the larger noise/signal ratio.

It is also easy to control the background electrolyte; not only its concentration (Fig.\ \ref{Fig3}), but also the ions that constitute the electrolyte.
While we always use Cl$^{-}$ for the anion in this study, we performed a series of calculations using different cations: Li$^{+}$, Na$^{+}$, K$^{+}$, and Cs$^{+}$ in increasing order in the ionic radius (Table \ref{tab:pauling}).
Because larger cations are less efficient in competing for space with the X$^{+}$ ions, they are more appropriate for the purpose of the cation of the background electrolyte in the sensor (Fig.\ \ref{Fig6}B).
Targeting even larger, possibly organic, molecules can result is even better efficiency.
Simulating such particles, however, might require special techniques.
In practice, these may stick to the nanopore wall and be less effective.

To assess what kind of analytes can be detected with such a device more efficiently, we changed both the size and the charge of the X ions (Fig.\ \ref{Fig6}C and D).

Our results show that it is easier to detect analyte ions with larger charge (Fig.\ \ref{Fig6}D). 
This corresponds to our intuition because X ions with higher valences, $z_{\mathrm{X}}$, repulse the K$^{+}$ ions more strongly, therefore, they compete with them more efficiently.
The real reason, however, is that higher-valence X ions are attracted to the pore charge more strongly than K$^{+}$ ions, so they compete with K$^{+}$ ions more efficiently.
Also, their contribution to the total current is larger.

The effect of the radius of the X$^{+}$ ion, $R_{\mathrm{X}}$, however, is more counterintuitive.
The calculations show that it is easier to detect small X$^{+}$ ions (Fig.\ \ref{Fig6}C).
One might expect that large X$^{+}$ ions would decrease the effective cross section of the pore more effectively.
This way of (mechanical) thinking, however, is misleading because what matters is the outcome of the (thermodynamic) competition of the X$^{+}$ and K$^{+}$ ions for space in the pore.
In this competition, on the other hand, the small X$^{+}$ ions are more efficient than large ones.
This is in agreement with the previous result that it is better to use larger background cations.
What matters is the ratio of $R_{\mathrm{X}}/R_{+}$, and it is better if this ratio is smaller.

\section{Summary}
\label{sec:discuss}

We studied a reduced model for a nanopore sensor based on the selective competition of the analyte ion with the cation of the background electrolyte for the presence in the pore.
Although the model contains only a limited number of parameters, it still exhibits a complex behavior in which every parameter has a considerable influence on the response of the device given to the presence of the analyte ions in the bulk, which can be considered as the input signal of the device.
Our results show that (up to a reasonable signal to noise ratio) using low background electrolyte concentrations and low voltage in a narrow pore produce the largest sensor signals. 
Also, the most easily detected analytes are small, high valence ions. 

These results also indicate the parameter range where one must look for candidates for analyte ions and molecules that bind them.  
The analyte molecules should be small, for example. 
This narrows down the possible choices and finding a properly selective binding site is also a challenge.
When large molecules are to be detected, binding/unbinding events are so rare that chemical analysis should use single molecule methods rather than an ensemble method described here.
This is a start-up study that can be expanded in various directions such as using Brownian Dynamics to examine time dependence or finding specific analyte molecule--binding site pairs and modeling them with more realistic force fields.  

\begin{table}[t]
\caption{Summary of the dependence of the efficiency on various parameters.
Parameters are divided into two groups depending whether their increase or decrease improves efficiency.  
}
\label{tab:conclusion}
\renewcommand{\arraystretch}{0.75}
  \begin{tabular}{l| c c}
\toprule
  \multicolumn{3}{c}{Sensitivity can be improved by} \\ \hline
   Parameters & increasing & decreasing \\ \hline
   model	  & $\epsilon_{\mathrm{SW}}$, $d_{\mathrm{SW}}$, $N_{\mathrm{ring}}$& $R_{\mathrm{pore}}$ \\
   analyte  & $z_{\mathrm{X}}$	& $R_{\mathrm{X}}$ \\
   external & $R_{+}$ 	& $c_{+}$, $U$ \\ \toprule
  \end{tabular}
\end{table}

The merit of this study is that, despite its simplicity, the model includes very different forces such as local selective short-range interactions vs.\ long range Coulomb interactions, as well as a confined high-concentration pore region, where the finite size of ions matter, and  large low-concentration baths. 
Our method is able to meet these diverse requirements because it can efficiently handle current calculation, low concentrations, and complex three-dimensional modeling at the same time.
Every other method mentioned in Section \ref{sec:method} fails with one of these requirements.
The Poisson-Nernst-Planck theory, for example, cannot handle ionic correlations such as finite size effects.
Brownian Dynamics, on the other hand, has problems with low concentrations and with sampling small currents.
Density functional theory is difficult to implement in three dimensions.

The results show that it is possible to have a nanopore sensor model that can detect small analyte concentrations.
The sensitivity analysis performed in this work is summarized in Table \ref{tab:conclusion}.
How these findings can be turned into real devices will require more work both from modelers and experimentalists, but our feasibility study indicates that this is worthwhile.

\section*{Acknowledgments}

The financial support of the National Research, Development and Innovation Office -- NKFIH NN113527 in the framework of ERA Chemistry is gratefully acknowledged. 
Present article was published in the frame of the project GINOP-2.3.2-15-2016-00053. 
Supported by the UNKP-17-4 (to MV) and UNKP-17-1 (to EM) New National Excellence Program of the Ministry of Human Capacities.
We thank Dirk Gillespie for the helpful discussions.


\end{document}